\documentclass[12pt]{iopart}
\usepackage{graphicx}

\begin{document}
\title[AIRAPT 2003]{Unconventional superconductivity and normal state properties of $\varepsilon$-iron at high pressure}
\author{A T Holmes$^1$\footnote[1]{To whom correspondence should be addressed
(alexander.holmes@physics.unige.ch)}, D Jaccard$^1$, G Behr$^2$, Y
Inada$^3$ and Y Onuki$^3$}

\address{$^1$DPMC, University of Geneva, 24 Quai Ernest-Ansermet, CH1211 Geneva 4, Switzerland}
\address{$^2$IFW Dresden, P.O. Box 27 00 16, D-01171 Dresden, Germany}
\address{$^3$Graduate School of Science, Osaka University, Toyonaka, Osaka 560-0043, Japan}
\begin{center}
\today
\end{center}
\begin{abstract}
Following the discovery of superconductivity in
$\varepsilon$-iron, subsequent experiments hinted at non-Fermi
liquid behaviour of the normal phase and sensitive dependence of
the superconducting state on disorder, both signatures of
unconventional pairing. We report further resistive measurements
under pressure of samples of iron from multiple sources. The
normal state resistivity of $\varepsilon$-iron varied as
$\rho_0+AT^{5/3}$ at low temperature over the entire
superconducting pressure domain. The superconductivity could be
destroyed by mechanical work, and was restored by annealing,
demonstrating sensitivity to the residual resistivity $\rho_0$.
There is a strong correlation between the $\rho_0$ and $A$
coefficients and the superconducting critical temperature $T_c$.
Within the partial resistive transition there was a significant
current dependence, with $V(I)=a(I-I_0)+bI^2$, with $a\gg b$,
possibly indicating flux-flow resistivity, even in the absence of
an externally applied magnetic field.
\end{abstract}

\section{Introduction}

Since the discovery of superconductivity in the $\varepsilon$
phase of iron under pressure \cite{Shimizu}, several observations
have hinted that the electron pairing has an unconventional
origin. Firstly, the restricted pressure range of the
superconducting (SC) state is hard to explain by BCS theory
\cite{Mazin, Jarlborg1, Jarlborg2,Bose}. The partial resistive
transitions observed implied an unusual sensitivity to disorder, a
characteristic property of certain unconventional superconductors.
Proximity to the ferromagnetic phase suggests that spin
fluctuations could be involved in the relevant electronic
interactions.

Subsequent experiments confirmed the unusual properties of
$\varepsilon$-Fe. An attempt to observe superconductivity in a
high purity commercial sample failed \emph{[sample \#0, Goodfellow
4N+, with impurities (ppm): Ag $<$ 1; Al $<$ 1;Cu $<$ 1; Si 1]}.
However, to achieve the dimensions necessary for a pressure
experiment, the sample had been rolled down to a thickness of
around 3$\mu$m, introducing disorder which significantly raised
its residual resistivity.

Further experiments succeeded in obtaining complete
superconducting resistive transitions \cite{PLA} when special care
was taken over the choice of samples and their preparation, to
avoid not only chemical impurities but structural disorder due to
mechanical work.

Other indications about the nature of the superconductivity come
from the normal state resistivity of the $\varepsilon$ phase at
low temperature. This appeared, at a pressure $P=22.2\:$GPa, to
follow a $T^{5/3}$ power law with a strongly enhanced prefactor
over a broad range of temperature, a $T$-dependence expected of a
nearly ferromagnetic Fermi-liquid (NFFL). The presence of
ferromagnetic fluctuations could imply $p$-wave triplet
superconductivity, a scenario suggested in the case of ZrZn$_2$
\cite{Pfleiderer}, which has many features in common with iron
(partial resistive transitions, non-Fermi liquid power law,
sensitivity to disorder). This stands in contrast with band
structure calculations \cite{Mazin,Jarlborg1,Jarlborg2,Bose,Cohen}
where antiferromagnetic fluctuations are expected to dominate the
ground state, and $p$-wave superconductivity is not favoured.

The nature of the pressure-driven bcc-hcp $\alpha$-$\varepsilon$
transition complicates the situation.  The transition is
martensitic  \cite{Wang}, and its pressure width (but not its
hysteretic behaviour) is highly dependent on the pressure
conditions \cite{Taylor}. In the quasi-hydrostatic conditions of
the steatite medium used in our experiments, where pressure
gradients are up to 5\%, this should lead to a transition spread
over at least 10 GPa, very similar to the SC pressure range. The
superconductivity may therefore be somehow linked with the
structural transition itself, rather than being an intrinsic
property of the $\varepsilon$ phase.  In this series of
experiments we have not addressed this question, though it may
turn out to be a crucial one. Future studies using different
pressure media should help to clarify the issue.

\section{Experimental methods and sample preparation}

The high pressure experiments were carried out using the same
Bridgman anvil technique used for Ref.~\cite{PLA}. The aim of
these follow-up measurements was to explore further the role of
structural disorder in the appearance or otherwise of
superconductivity, and to confirm the non-Fermi liquid $T^{5/3}$
behaviour of the normal state resistivity in the $\varepsilon$
phase.

We wished to investigate whether superconductivity could be
destroyed, and/or recovered or induced by mechanical work and
annealing, respectively. There were four samples in this study,
three came from the superconducting batches reported in
\cite{PLA}, one from Dresden (\#6) and two from Osaka (\#3,4). The
remaining sample (\#5) \emph{[Goodfellow 99.998\% Fe, impurities
(ppm): Ag $<$ 1; Al $<$ 1; Cu $<$ 1; Mg 1; Mn 1; Ni $<$ 1]} was
obtained from the same commercial metallurgical supplier as \#0.
No trace of superconductivity had been found in a previous
investigation of sample \#0, where the residual resistivity ratio
(RRR) was 15. (The RRR is defined by $\rho(298\:\rm
K)/\rho(4.2\:\rm K)$ and used as a measure of sample quality.)

The four samples were prepared as follows: \\
They were rolled down to a thickness appropriate for the pressure
experiment($\sim 15\:\mu\rm m$). Initial RRR values of up to 300
were reduced to $\sim$50 by this operation. Three of the samples
(\#4--6) were subsequently annealed in an induction furnace for 24
hours at 1000$^\circ\rm C$ in a high vacuum ($<10^{-7}\rm mbar$).
After annealing, RRRs in the range 245--310 were recovered in
samples from all sources, and grain sizes comparable to the
eventual sample length were observed. Sample (\#3) was left
unannealed.

The samples were then cut using a razor blade to the appropriate
width and length. They were arranged in the pressure cell, along
with a lead manometer, so that current passed through in series.
Four-point resistance measurements could be carried out separately
on each sample. A knowledge of the each sample's dimensions
enabled the absolute resistivity to be calculated with an error of
less than 15\% over the whole pressure range, neglecting the
compressibility.

\section{Experimental Results and Discussion}
\begin{figure}
\begin{center}
    \includegraphics[width=.6\columnwidth]{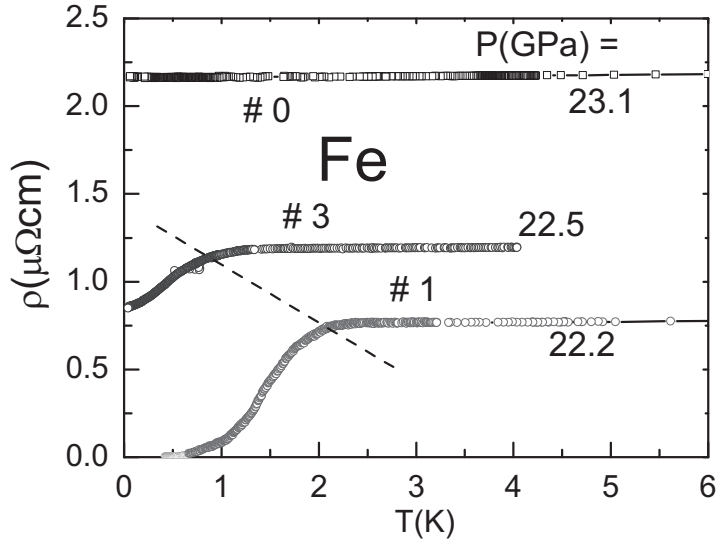}%
\caption{\label{fig:TcRho0}Effect of disorder in different samples
on the superconducting transition of iron. The dashed line
suggests the $T_{\rm c}$ variation vs. $\rho_0$.}
\end{center}
\end{figure}
The effect of disorder on the superconducting state can be seen in
Fig.~\ref{fig:TcRho0}, where the resistivity of three
representative samples \#0 \#1 and \#3 are shown at pressures
close to the maximum $T_{\rm c}(P)$; their residual resistivities
were 2.13, 0.77, and 1.19 $\mu\Omega \rm cm$ respectively. With
increasing disorder, measured by $\rho_0$, the resistive
transition becomes only partially complete, then disappears
entirely. The remaining samples had resistance drops somewhere
between \#1 and \#3. A transition in \#3, the unannealed sample,
was only detectable once the pressure had reached 22$\:$GPa, while
the annealed samples started to show signs of superconductivity
almost as soon as the structural transition had initiated (as
identified in the room-temperature resistivity). Rolling and
annealing thus have opposite effects, tending to suppress or
restore the superconductivity, with associated $\rho_0$ variation.
The metallurgical state, rather than the chemical impurity level
appears to be the dominant factor. Sample \#1, however, exhibited
the highest $T_{\rm c}$ ($2.5\:$K). It was not annealed or rolled,
but cut to shape with a diamond saw, implying that the annealing
process may introduce some contamination which can affect $T_{\rm
c}$. These results suggest that if $\rho_0$ can be lowered
further, $T_{\rm c}$ values larger than 2.5$\:$K could be
obtained.

\begin{figure}
\begin{center}
    \includegraphics[width=.6\columnwidth]{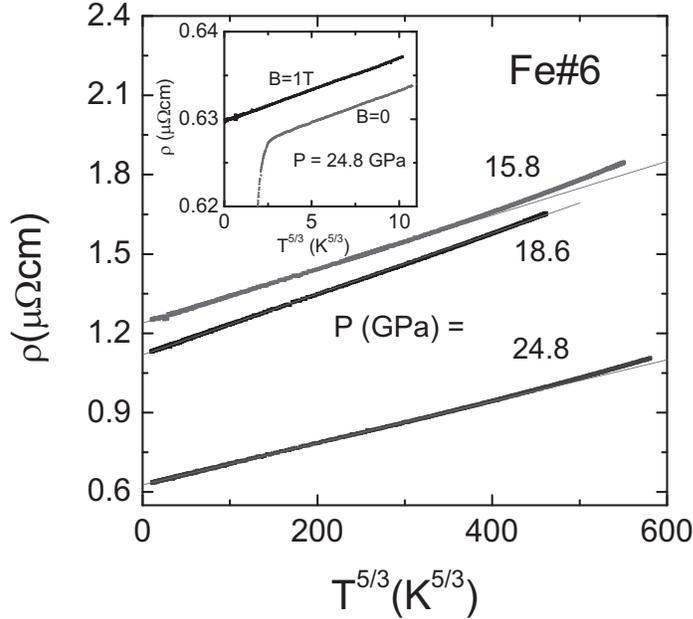}%
\caption{\label{fig:rhoT53}The resistivity of $\varepsilon$-Fe
 on a $T^{5/3}$ scale at selected pressures. Inset: in a magnetic field
$B$ = 1 T, the $T^{5/3}$ dependence persists below $T_{\rm c}$.}
\end{center}
\end{figure}

Figure \ref{fig:rhoT53} shows the relation $\rho=\rho_0+AT^{5/3}$
in sample \#6 at several pressures, extending up to nearly
35$\:K$, with the upward deviation at higher temperature ascribed
to electron-phonon scattering. Similar temperature dependence was
seen in the other samples. If the exponent was allowed to vary, a
fit of $\rho=\rho_0+\tilde A T^n$ gave $n$ between 1.66 and 1.75
over the entire superconducting pressure range. There is an
evident change with pressure of the residual resistivity $\rho_0$,
along with the temperature coefficient $A$.

The inset of Fig.~\ref{fig:rhoT53} shows that the $T^{5/3}$ law
remains valid down to very low temperature when the
superconductivity is suppressed by a magnetic field of 1$\:$T. The
normal-state resistance varied with magnetic field $B$ as
$(\rho-\rho_0)\propto B^{3/2}$ up to 8$\:$T at 4.2$\:$K and
700$\:$mK.
\begin{figure}
\begin{center}
    \includegraphics[width=.6\columnwidth]{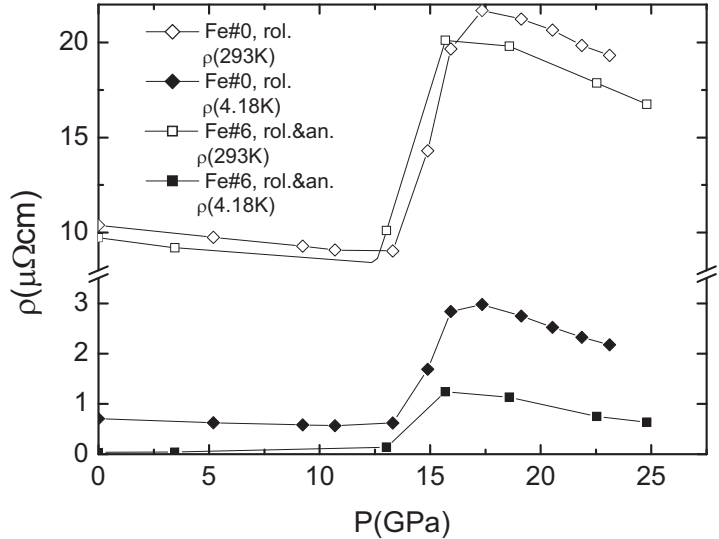}%
\caption{\label{fig:Behr_GF}Resistivity signature of the
$\alpha$-$\varepsilon$ transition of iron with increasing pressure
at 4.18 K and 293 K.}
\end{center}
\end{figure}

Figure \ref{fig:Behr_GF} shows the resistivity at room temperature
and 4.2$\:$K in a superconducting (\#6) and a non-superconducting
sample (\#0) as the pressure is increased. The
$\alpha$-$\varepsilon$ transition is clearly visible at both
temperatures, with a substantial increase in $\rho$. However, the
martensitic $\alpha$-$\varepsilon$ transition should start at $p <
10\:$GPa in a steatite medium according to M\"{o}ssbauer effect
experiments \cite{Taylor}. The kink observed in $\rho$ at higher
pressures, around 12$\:$GPa, with no precursor signs is therefore
unexpected. It could nevertheless be attributed to the
disconnection of the conducting $\alpha$ region by the growing
amount of $\varepsilon$ phase.

After a rounded maximum, the resistance at both temperatures drops
quickly and more or less linearly with pressure, and in both cases
can be extrapolated to the $\alpha$-Fe values at pressures close
to the disappearance of superconductivity.

The origin of the pressure-induced change in resistivity at room
temperature might be due to a change in the electron-phonon
coupling, or more likely to the large increase in spin disorder
when passing from the magnetically ordered $\alpha$-state to the
$\varepsilon$ phase. The effect of magnetic ordering on the
resistivity has been explored in metastable non-magnetic
$\gamma$-Fe at ambient pressure \cite{Bohnenkamp}, where a large
difference is found in the room-temperature resistivity between
the magnetically ordered and non-magnetic states. We might
therefore expect $\varepsilon$-Fe to be analogous.

At low temperature, where the phonon contribution is negligible,
the increase in $\rho$ is associated with additional disorder, but
the very similar pressure dependence of $\rho$ at these two
temperatures suggests that the scattering is at least in part of
the same magnetic origin.

The two samples shown in Fig.~\ref{fig:Behr_GF} behave similarly
at room temperature, apart from a shift ascribed to the residual
term.  At 4.2$\:$K, interestingly, the increase in resistivity
associated with the $\alpha-\varepsilon$ transition is smaller for
a lower $\rho_0$. Single crystalline iron whiskers, which can be
prepared with RRRs around 1000 \cite{Taylor68}, therefore offer a
good path to obtaining a higher $T_{\rm c}$.

\begin{figure}
\begin{center}
    \includegraphics[width=.6\columnwidth]{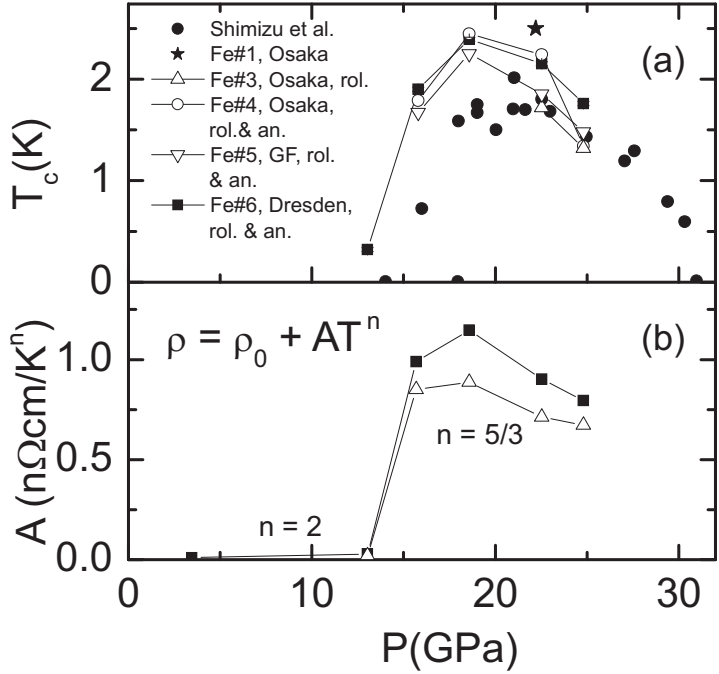}%
\caption{\label{fig:TcandA}Pressure dependence of (a) the
 superconducting transition of various Fe samples and (b) the $A$
coefficient in the law $\rho=\rho_0+AT^{5/3}$.}
\end{center}
\end{figure}

Figure \ref{fig:TcandA} shows $T_{\rm c}$ as a function of
pressure, compared with the $A$ coefficient of a fit to the
resistivity using a $T^2$ power law in the $\alpha$ phase and a
$T^{5/3}$ law in the $\varepsilon$ phase. The $T_{\rm c}(P)$ phase
diagram in Ref.~\cite{Shimizu} is confirmed by our measurements.
The superconductivity started to emerge exactly at the kink in
$\rho(P)$. 

As the resistive transitions were highly current dependent and
partial in most cases, the onset was used as a criteria for
$T_{\rm c}$, i.e. $T_{\rm c}$ was defined where the resistivity
curve started to deviate visibly from the normal state temperature
dependence. Surprisingly, the onset of superconductivity was found
to be much less sensitive to current than other criteria, such as
a 50\%, or even 1\% drop in resistance from the normal state. The
resistivity drop became larger in every sample as the pressure
increased, even beyond the maximum in $T_{\rm c}$.

The very large value of $A$ in the SC domain
[Fig.~\ref{fig:TcandA}(b)] is evidence of a strongly correlated
electronic phase. The value of $A$ reflects the strength of
interaction between the electrons and spin fluctuations, thus the
correlation seen between $A(P)$ and $T_{\rm c}(P)$ is strong
evidence for a magnetically mediated pairing scenario. Moreover,
there appears to be a positive correlation between $A$ and $T_{\rm
c}$ in different samples (including those not shown). It is
somewhat surprising that the value of $A$ would vary between
samples, but as they were all measured in the same cell, and their
resistivities normalised at ambient pressure, we can have some
confidence that this is a genuine observation.

Spin-fluctuation mediated superconductivity is usually associated
with a well defined quantum critical point at a particular
critical pressure $P_c$, where non-Fermi liquid behavior of the
resistivity is expected \cite{Millis}. In contrast, a $T^2$
dependence was never recovered in the entire SC domain in our
measurements.

\begin{figure}
\begin{center}
    \includegraphics[width=.6\columnwidth]{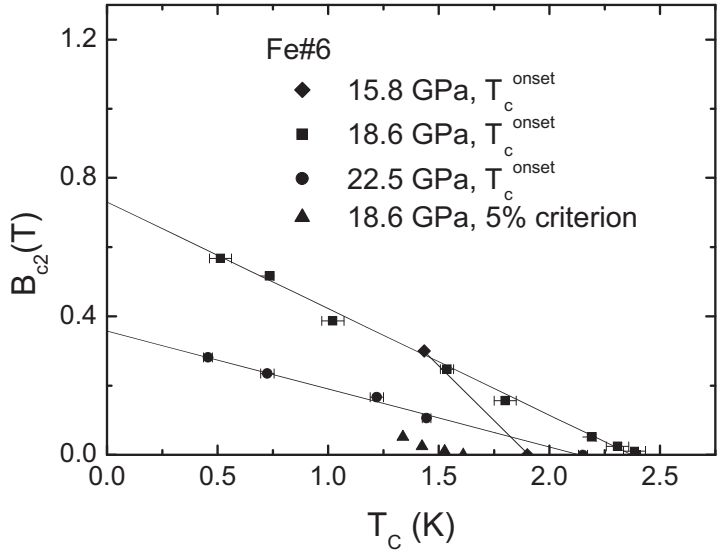}%
\caption{\label{fig:Hc2}The critical field of $\varepsilon$-Fe at
selected pressures. Two $T_{\rm c}$ criteria are used at 18.6
GPa.}
\end{center}
\end{figure}

Figure \ref{fig:Hc2} shows the upper critical field $B_{\rm c2}$
for sample \#6 determined using an onset criterion for $T_{\rm
c}$. The extrapolated value of $B_{\rm c2}(T=0)$ (0.73$\:$T at
18.6$\:$GPa and $\sim 0.35\:$T at 22.5$\:$GPa) is much larger(by a
factor of up to 70) than that of the lead manometer, which has a
nearly identical $T_{\rm c}$. $B_{\rm c2}$ is also linear, without
the usual curvature, at least down to $T_{\rm c}/5$.
Interpretation of this is difficult, because the internal magnetic
field in the sample itself may be higher still. Hints that this is
the case come from the current dependence of the resistance within
the transition. This could be fitted very well to a flux-flow type
linear current-voltage relation, accompanied by a small quadratic
term, i.e. $V=a(I-I_0)+bI^2$ where $a\gg b$, and $I_0$ was very
small (an equivalent current density of 0.17~$\rm A cm^{-2}$),
indicating an extremely low pinning density. This flux-flow like
behaviour was present even with no externally applied magnetic
field.

Taking a value of 0.73$\:$T for $B_{\rm c2}(T=0)$, the coherence
length $\xi$ is around 20$\:$nm, which is comparable to the mean
free path $l$, derived from band structure calculations
\cite{Private}.  A requirement for the clean limit, i.e.
$\xi\leq\l$, is another feature of unconventional
superconductivity.

The initial slope $B_{\rm c2}'=\partial B_{\rm c2}/\partial T_{\rm
c}$, which is proportional to the electronic effective mass in a
clean limit scenario, appears to be larger for a $p$ below the
$T_{\rm c}$ maximum, suggesting a critical region close to the
emergence of superconductivity. In any case, $B_{\rm c2}'$ reaches
values larger than that observed in any other SC element in the
periodic table.

\section{Conclusions}
In a narrow pressure window, both the superconductivity and normal
phase of iron show features which point to a key role for spin
fluctuations.  The closest similar example seems to be ZrZn$_2$
($T_{\rm c}\sim0.3\:$K)  \cite{Pfleiderer}.  The higher $T_{\rm
c}$ of iron may result from stronger magnetic interactions,
reflected for example in the large Curie temperature of iron's
ferromagnetic phase. Experimentally, it is not clear whether
superconductivity is an intrinsic property of $\varepsilon$-Fe or
is related to the martensitic transition.

Iron is the first simple element to be observed exhibiting
non-Fermi liquid behaviour in its resistivity, and to be a
candidate for spin-mediated superconductivity.

\end{document}